# Timbre Space Representation of a Subtractive Synthesizer

Cyrus Vahidi[1,†], George Fazekas[1], Charalampos Saitis[1] and Alessandro Palladini[2]
[1] Centre for Digital Music, Queen Mary University of London, London, United Kingdom
[2] Music Tribe Research, Manchester, United Kingdom
[†] Corresponding author: c.vahidi@qmul.ac.uk

## Introduction

Sound synthesis modules include oscillators, filters and amplifiers, that can be routed to form a subtractive synthesizer (Moog, 1964). Sound synthesizers provide a rich, abstract timbre palette, that is not achievable under the physical constraints of acoustic musical instruments. Synthesis modules are controlled by low-level signal-processing parameters; this parameterization does not allow control of perceptual timbre dimensions and requires expert knowledge to smoothly navigate the space of timbres. Acoustic descriptions of perceptual timbre spaces have been discovered, with multidimensional scaling (MDS) of pairwise dissimilarity ratings, obtained for real and synthetic orchestral musical instrument stimuli (McAdams et al., 2019). Timbre space studies have been conducted jointly between acoustic and synthesized sounds, where the synthesized stimuli has clear definition (Zacharakis et al., 2015). In contrast to instrument sounds, synthesized sounds are not generated by a physical body; familiarity with the object may have less influence on the dissimilarity judgements of listeners (Siedenburg et al., 2016).

Abstractions for interaction with sound synthesis have been advanced with deep neural networks. These generative systems are capable of high-level characterization of timbre as a control structure (Esling et al., 2020). It has been shown that the alignment of generative synthesis spaces with timbre dissimilarity distances, produces sparser latent representation spaces (Esling et al., 2018). We aim to generate new domain knowledge for generative modelling of sound synthesizers.

In this study, we produce a geometrically scaled perceptual timbre space from dissimilarity ratings of subtractive synthesized sounds and correlate the resulting dimensions with a set of acoustic descriptors. We curate a set of 15 sounds, produced by a synthesis model that uses varying source waveforms, frequency modulation (FM) and a lowpass filter with an enveloped cutoff frequency. Pairwise dissimilarity ratings were collected within an online browser-based experiment. We hypothesized that a varied waveform input source and enveloped filter would act as the main vehicles for timbral variation, providing novel acoustic correlates for the perception of synthesized timbres.

## Method

*Participants*
35 participants (age 28.2 mean, 5.8 std, 21-46) for this experiment consisted of domain experts and musically critical listeners. A total of 41 responses were received. Participants who were not included in the final analysis either reported hearing issues or excessively violated the experiment control. Participants were recruited through the Centre for Digital Music and music informatics mailing lists.

*Stimuli and Presentation*
Figure 1 shows a block diagram of the digital subtractive synthesizer which was implemented in SuperCollider to generate the set of 15 stimuli. All oscillators were fixed at 440 Hz (A4) with an initial phase of 0. FM was applied with a modulator frequency that is an integer multiple of the carrier. Tones were normalized for A-weighted RMS sound level and constrained to 1000ms in duration. The stimuli were composed from either of two oscillators, a pulse wave or a sawtooth. The source was optionally frequency modulated and then passed into a resonant low pass filter. The lowpass filter's cutoff frequency, $\omega_c$, was time-varied by an attack-decay-sustain-release (ADSR) envelope. The master gain, $g$, was also time-varied with a separate ADSR.

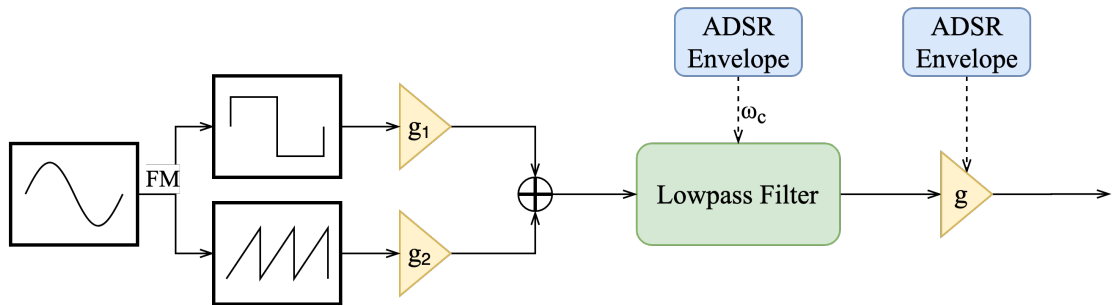

*Figure 1: Block diagram of a subtractive synthesizer.*

*Procedure*
The experimental procedure was approved by Queen Mary University of London Ethics of Research Committee (ref QMREC2445).

Conducting a listening experiment in the browser presents challenges, such as varying listening environments. In order to mitigate the effects of insufficient audition quality, we used a headphone screening task (Woods et al., 2017) to ensure greater control over stimuli presentation.

Each participant provided a single rating for each sound pair A-B. The direction of presentation, A-B vs B-A, was selected randomly for each pair and each participant. This resulted in a total of 120 pairwise ratings per participant. Participants were instructed to provide a dissimilarity rating between 0 (identical) and 9 (very dissimilar), via a slider that was discretized in steps of 0.5. As a control, participants were instructed to provide a 0 rating for identical sounds. Participants were also given the opportunity to familiarize with the sound set and task. A pair of sounds could be replayed infinitely before submitting a rating.

*Non-metric Multidimensional Scaling*
Multidimensional scaling (MDS) is a statistical dimensionality reduction technique that aims to preserve the geometric relationships between data objects (Kruskal, 1964a). The dissimilarity data were analysed by means of MDS, and the dimensions of the timbre space were examined in terms of audio content descriptors that were computed from the acoustic signals. Non-metric multidimensional scaling (Kruskal, 1964b), provided by MATLAB, was applied to the mean dissimilarity matrix across participants.

*Feature Selection*
Suitable acoustic descriptors were identified from literature on acoustic feature extraction (Peeters et al., 2004, 2011) and extracted using Essentia (Bogdanov et al., 2013). A Hann window and hop length of 2048 and 512 samples, at a sampling rate of 44.1 kHz, were used for the magnitude Short-time Fourier transform (STFT) spectral feature extraction. The mean statistic was computed across time frames. A Spearman correlation matrix was computed for the initial set of acoustic descriptors, discarding features that showed a strong collinearity with others ($r > 0.8$), for example spectral centroid and spectral roll-off[1].

## Results

Figure 2 shows the metrics for MDS solutions of varying dimensions. *Stress-1* and $R^2$ goodness-of-fit metrics were computed between the original dissimilarities and new MDS disparities. There was a significant improvement in fit when moving from 3 to 4 dimensions and greater interpretability of the resulting acoustic correlations. Hence, a four-dimensional MDS solution was used for analysis (*Stress-1 = .047, $R^2$ = .91)*.

---

[1] Collinear features were eliminated based on their relevance in timbre literature and interpretability.

Table 1 shows the Pearson correlation matrix, computed between the final set of acoustic content descriptors, and the 4 perceptual dimensions derived from MDS analysis. *Dimension 1* indicated significant correlations with *spectral complexity* ($p < 0.01$), *spectral flux* ($p < 0.01$), *log-attack time* ($p < 0.05$) and *tristimulus 3* ($p < 0.01$). *Dimension 2* indicated a significant positive correlation with *spectral decrease* ($p < 0.05$). *Dimension 3* demonstrated significant positive correlations with *spectral kurtosis* ($p$

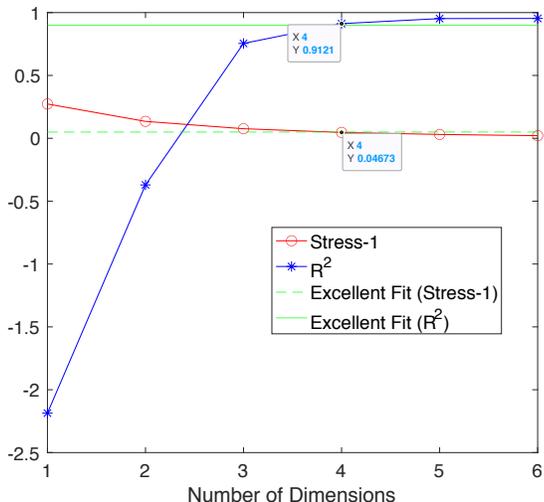
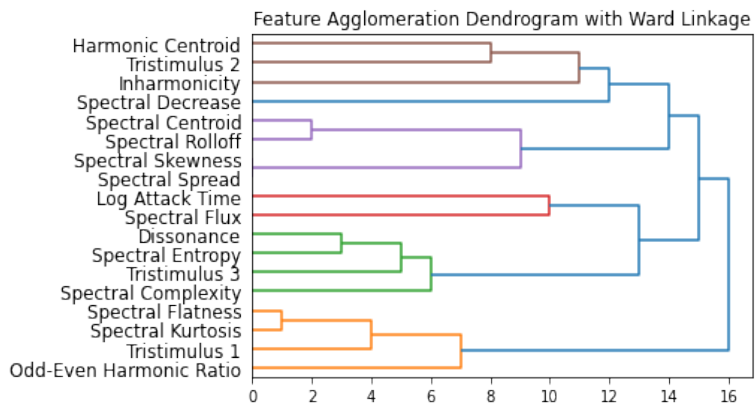

*Figure 3: Stress-1 and $R^2$ metrics*  *Figure 2: Feature Agglomeration Dendrogram*

$< 0.05$) and *odd-even harmonic ratio* ($p < 0.05$), and negative correlations with *tristimulus 2* ($p < 0.05$) and *spectral centroid* ($p < 0.05$). *Dimension 4* demonstrated a significant negative correlation with *spectral centroid* ($p < 0.01$).

|  | MDS Dimension | | | |
|---|---|---|---|---|
| *Acoustic Descriptor* | 1 | 2 | 3 | 4 |
| *Spectral Complexity* | **.75**** | .39 | -.23 | .10 |
| *Spectral Flux* | **.68**** | -.24 | -.08 | .41 |
| *Log Attack Time* | **.60*** | -.39 | -.26 | -.37 |
| *Tristimulus 3* | **.75**** | .31 | -.23 | -.37 |
| *Spectral Decrease* | -.23 | **.58*** | .24 | -.43 |
| *Tristimulus 2* | -.44 | .14 | **-.51*** | .30 |
| *Spectral Kurtosis* | -.01 | -.43 | **.61*** | .35 |
| *Odd-Even Ratio* | -.34 | -.30 | **.56*** | .14 |
| *Spectral Centroid* | .05 | .28 | **-.52*** | **-.65**** |

*Table 1: Acoustic Feature vs Perceptual Dimension Pearson Correlations. \* - $p < 0.05$, \*\* - $p < 0.01$*

## Discussion

The results indicate some anticipated acoustical correlates in the context of the literature, as well as additional dimensions that can be explained by the design of the synthesis model.

The first perceptual dimension appears to be explained by the content of spectral peaks and their change over time. Note that spectral flux and log-attack are entangled. A plausible explanation for this is the design of the sound set. The attack of several sounds is implicit in the rising cutoff of the envelope filter, that moves from 0 Hz to N Hz over a time determined by the filter envelope's attack time. Hence, we define *dimension 1* to relate to spectrotemporal variation and harmonic peaks.

*Dimension 2* can be interpreted to relate to the varying cutoff of the lowpass filter, as the spectral decrease indicates the ratio between low and high frequency components. *Dimension 3's* acoustical correlates can be interpreted to be related to the (a) strength of the FM modulation index, which when increased reduces flatness around the centroid, and (b) the input waveforms. This accounts for spectral detail and distribution of harmonic partials. *Dimension 4's* significant correlation with the spectral centroid is an anticipated result, given the ubiquity of the spectral centroid in previous perceptual timbre space studies, and the effects of harmonic FM and resonant filters on shaping timbral brightness.

## Conclusions and Future Work

In the present study, additional dimensions explaining perception of synthesized timbres were observed, relative to existing timbre studies on acoustic and synthesized sounds. A study with a larger, more diverse sound set is necessary to disentangle log-attack time and spectral flux. The results of this study shall contribute to future large-scale data generation and perceptual domain knowledge for generative modelling of sound synthesizer timbre spaces. With these results, we can move further towards perceptual representation of synthesized timbres.

## Acknowledgements

We acknowledge Benjamin Hayes for his development of the framework used for online dissimilarity data collection. We thank all who participated in the listening experiment. The author is a research student at the UKRI CDT in AI and Music, supported jointly by the UKRI [grant number EP/S022694/1] and Music Tribe.